\documentstyle[pra,aps]{revtex}

\tightenlines
\draft 
\begin{document}
\draft

\title{Simulating entangled sources by classically correlated sources and quantum
imaging}
\author{Morton H. Rubin}
\address{Department of Physics \\
University of Maryland, Baltimore County \\
Baltimore, MD 21228-5398}
\date{\today }
\maketitle

\begin{abstract}
We discuss the problem of when a set of measurements made on an entangled
source can be simulated with a classically correlated source. This is
discussed in general and some examples are given. The question of which
aspects of quantum imaging can be simulated by classically correlated
sources is examined.
\end{abstract}


\section{Introduction}

Recently there has been a discussion in the literature about when an imaging
experiment carried out with spontaneous parametric down conversion (SPDC)
can be simulated classically \cite{Bennink,Abouraddy}. We wish to discuss
this problem in the general setting of quantum information theory and to
give a class of experiments in which the measurements with a given entangled
source can be reproduced using a classically correlated source. In addition,
we show that certain aspects of the ''ghost'' interference experiment \cite
{Strekalov} can be reproduced using a classically correlated source. It is
very important to have a clear understanding of what is being simulated. We
show that while certain results can be reproduced by classically correlated
sources, those results that depend on the phase of the two photon amplitude
or biphoton can not. For a review of quantum imaging see ref. \cite
{Shih,Lugiato}.

There has been a great deal of work done on determining whether a given
state is entangled. The problem discussed here is different, we ask when the
results of the experiment using a source that produces an entangled state
can be reproduced by replacing the entangled state by a separable state.  In
the imaging experiments an entangled initial state is generated by SPDC and
is transformed into the output state by the fields propagating through a
linear optical system. The most important point is that the system has loss
in it, indeed, the image is formed by removing photons from one of the
beams. Loss may transform the state from an entangled state to a classically
correlated state. When this happens all the measurements can be reproduced
by a classically correlated source. However, this is not the general case
since, even after losses, the state may remain entangled. Consequently, only
certain measurements may be reproduced classically. This is the case we wish
to discuss in this paper. The problem of substituting a stochastic classical
source for a quantum source in optics experiments has been discussed many
times. In the context of this paper, two reviews by Belinskii and Klyshko 
\cite{Belinskii} are of interest.

It is important to be clear what we mean by an experiment. The original
imaging experiment using SPDC \cite{Pittman} consisted of a source of
entangled photon pairs with each photon passing through a linear optical
system to detectors. A measurement consisted of the coincident detection of
the pair.  Each photon was detected locally and the coincidence triggering
corresponded to classical communication. The outcome was summed over many
such pairs. In the original experiment one detector scanned the image plane
but this can be eliminated by placing a CCD detector in the image plane. In
this case the experiment is series of measurements performed on a fixed
setup that measures the intensity correlation function $\langle I_{A}({\bf r}%
_{1})I_{B}({\bf r}_{2j})\rangle $ where the $I$ 's are the normally ordered
intensities and $j=1,\cdots ,N$ indexes the detectors in the CCD array. Note
that these are $N$ commuting operators. In the experiment of Bennink, {\it %
et. al}. \cite{Bennink} the source is replaced by a laser, beam splitters, a
chopper and a moving mirror. In this case each measurement corresponds to a
point by point scanning of the object plane. Light detected behind the
object triggers the image plane detectors to their {\it on state} and the
detector corresponding to the image point responds. The two experiments are
related by replacing the SPDC source by a classical source. The classical
source is meant to simulate a stochastic source that has a large bandwidth
in the transverse wave numbers. If this bandwidth is the same as the
bandwidth for SPDC, there will be no image behind the object plane.

These experiments should be contrasted with a test of a Bell inequality \cite
{Bell}. Such a test consists of a series of experiments for which the source
is fixed but the measurement operators are different. This is important
because the quantities measured at {\it each} site are non-commuting
operators, so they can not be measured simultaneously. It has been
recognized for a long time that any experiment which only measures a fixed
set of local operators can be reproduced by a local hidden variable theory
in which the probabilities for each outcome of the experiment are themselves
taken as the local variables. Then any correlations that exist can be
modelled by separable density matrices, a simple example is given in section
III. The experimental results of \cite{Bennink} are consistent with this
conclusion. Some SPDC experiments can be understood in terms of a picture
first proposed by Klyshko \cite{Klyshko,Pittman1} in which one detector is
viewed as a source of photons propagating backwards in time, reflecting off
the SPDC crystal, and then propagating forward in time to the second
detector. It is often not appreciated that this picture has a rigorous
mathematical foundation for maximally entangled photon pairs. This picture
is genralized in \ref{KlyshkoPicture} so it can be used to explain the
experiment similar to that in ref. \cite{Bennink}. However, as can be seen,
results that depend on fixed phase relations between two-photon amplitudes
can not be simulated. Furthermore, it should be remembered that the quantum
or classical nature of a state is determined by the statistics of the
experiment, so even if the average of a set of measurements can be
reproduced classically, it does not follow that the fluctuations in
correlation measurements can be reproduced by classically correlated
systems. In the imaging experiment there are two important results such
that, as we shall see below, one can be simulated but the other can not. For
a different perspective on this problem see ref. \cite{DAngelo}.

In the following we will analyze the statements made above and give examples
of when an experiment using an entangled input state can be simulated by a
classically correlated input state. We first review some well-known results
about measurements and entanglement, and apply them to the simple case of
two qubits to illustrate some of the issues discussed above. Then we discuss
the imaging experiments.

\section{Background}

For all the experiments that we discuss assume that the source produces a
pair of subsystems $A$ and $B$ in a state $\rho _{AB}.$ The subsystems are
sent to two separate sites, also denoted by A and B, where local
measurements are made. The Hilbert space describing the system is $%
H=H_{A}\otimes H_{B}.$ The set of observables is defined by $%
\{O_{AB}^{(i,j)}=O_{A}^{(i)}\otimes O_{B}^{(j)}\quad (i,j)\in S\}$ acting on 
$H$ where the sets of Hermitian operators $\{O_{A}^{(i)}\}$ and $%
\{O_{B}^{(j)}\}$ act on $H_{A}$ and $H_{B,}$ respectively. A state $\sigma
_{AB}$ is said to be separable or classically correlated if it can be
written in the form 
\begin{equation}
\sigma _{AB}=\sum_{u}p_{u}\sigma _{A}^{(u)}\otimes \sigma _{B}^{(u)}
\label{seprable}
\end{equation}
where $\{p_{u}\}$ is a probability distribution, {\it i.e. }a set of
positive numbers that sums to $1,$ and $\sigma _{A}$ ($\sigma _{B}$) is a
density matrix acting on $H_{A}$ ($H_{B}$)$.$ An experiment will be defined
to be classical (more precisely, classically correlated) if there exists a
separable state such that for all $(i,j)\in S$ 
\begin{equation}
trO_{AB}^{(i,j)}\rho _{AB}=trO_{AB}^{(i,j)}\sigma _{AB}  \label{requirement}
\end{equation}
We may actually be content to simulate one observable that is a linear
combination of direct product observables, 
\begin{equation}
O_{AB}=\sum_{ij}q_{ij}O_{AB}^{(i,j)}.  \label{ZO}
\end{equation}
All of these are correlation experiments for which the experimentalists
agree which pair of observables they measure for a given realization of the
experiment. For example if the polarization of a pair of photons is measured
then $i=H$ or $V$ corresponding to the measurement of a horizontal or
vertical polarization in A, and similarly for the measurement in B, $j=H$ or 
$V$. The four possible projection operators are not only pairwise commuting
but are locally pairwise commuting, consequently, the experiment can be
performed using one fixed set-up in each laboratory. On the other hand if,
for example, $i=H,V,R$ and $L$ where $R$ and $L$ refer to right and left
hand polarized light, then the observables in A do not all commute and can
not be measured simultaneously\footnote{%
Of course it is possible to use a beam splitter to send half the photons to
the pair of detectors that measures H and V and the other half to detectors
that measure R and L but this does not change the argument.}. There may be
more than one classically correlated state that can simulate a given
experiment. If this is the case, the set of such states will form a convex
set \cite{convexset}. A simple illustrative example for qubits is given in
section III where these measurements are discussed in a geometrical context.

We illustrate how an experiment on an entangled state can be simulated using
a separable state by a simple example. Consider an entangled pure state $%
\Psi $, let $O_{AB}=O_{A}\otimes O_{B}$ be an observable, and let $\left\{
\phi _{j}\right\} $ be the set of eigenvectors of $O_{A}$. Let $\left\{ \chi
_{j}\right\} $ be an orthonormal basis of $H_{B}$ and consider the arbitrary
state 
\begin{equation}
\Psi =\sum_{j,k=1}^{N}c_{jk}\phi _{j}\otimes \chi _{k}.  \label{SchmidtRep}
\end{equation}
The measurement of $O_{AB}$ gives 
\begin{equation}
tr\left( O_{AB}|\Psi \rangle \langle \Psi |\right) =\sum_{j}\lambda
_{j}\langle \chi _{j}|O_{B}|\chi _{j}\rangle |c_{j}|^{2}  \label{Zshow1}
\end{equation}
where $O_{A}\phi _{i}=\lambda _{i}\phi _{i}.$ It is not difficult to see
that we get the same result for the classically correlated state 
\begin{equation}
\sigma =\sum_{j}|\phi _{j}\rangle \langle \phi _{j}|\otimes |\omega
_{j}\rangle \langle \omega _{j}|\quad |\omega _{j}\rangle
=\sum_{k}c_{jk}|\chi _{k}\rangle .\quad   \label{Zshow2}
\end{equation}
Therefore, any simple local measurement, that is, a measurement of a direct
product of local operators, can be simulated by a separable measurement.

Now consider a different observable obtained by a unitary transformation
applied to $O_{A}$, $O_{AB}^{\prime }=O_{A}^{\prime }\otimes O_{B},$ such
that $O_{A}^{\prime }=$ $U^{\dagger }O_{A}U$ and $O_{A}$ do not commute. For
example if subsystem $A$ is a spin-1/2 particle and $O_{A}=\sigma _{z}$ take 
$O_{A}^{\prime }=\sigma _{x}.$ It is not difficult to show that the new
measurements are not produced by the same classically correlated state since 
$\langle \phi _{k}|O_{A}^{\prime }|\phi _{j}\rangle \neq \lambda _{j}\delta
_{kj}$ in general. Finally if $\{O_{AB}^{(i,j)},$ $i,j\in S\}$ where the $%
\{O_{A}^{(i)}\}$ are pairwise commuting operators, this result immediately
generalizes since the $\{\phi _{j}\}$ can be taken as simultaneous
eigenvectors of all the $O_{A}^{(i)}.$

It is worth noting that if $trO_{AB}=0$ we can always find a separable state 
$\sigma (s_{0})$ such that $trO_{AB}\rho =(1/s_{0})trO_{AB}\sigma (s_{0}).$
To see this let $\rho $ be the given entangled state on an $N$ dimensional
Hilbert space, $I_{N}$ the identity operator, and $D_{0}$ be the completely
random state, $D_{0}=(1/N)I_{N}$ \cite{random}. The convex combination $%
\sigma (s)=(1-s)D_{0}+s\rho $ is a state of the system which is entangled
for $s=1$ and classically correlated for $s=0.$ Therefore, there exists an $%
s_{0}\in (0,1)$ such that the state $\sigma (s)=(1-s)D_{0}+s\rho $ is
classically correlated for $s\leq s_{0}$ \cite{Tarrach}. The reason for
mentioning this is that many experiments do not look at absolute values but
at relative values for which the normalization of the density matrices is
not important.

In a typical experiment, the state of the system measured, $\rho _{AB},$ is
related to an input state, $\rho _{0AB},$ by a completely positive, trace
preserving, linear transformation 
\begin{equation}
\rho _{AB}=T(\rho _{0AB}).  \label{PositiveMap}
\end{equation}
The Kraus representation of $T$ \cite{Kraus,Peres} is 
\begin{equation}
\rho _{AB}=T(\rho _{0AB})=\sum_{k=1}^{n}V_{k}\rho _{0AB}V_{k}^{\dagger }
\label{KrausRep}
\end{equation}
where the set of operators $\{V_{k},k=1,\cdots ,n\}$ acting on $H$ satisfy 
\begin{equation}
\sum_{k=1}^{n}V_{k}^{\dagger }V_{k}=I_{H}  \label{normalization}
\end{equation}
with $I_{H}$ is the identity operator on $H.$ In the experiments of interest
in this paper, each of the operators $V_{k}$ is a direct product of
operators acting on $H_{A}$ and $H_{B}$, 
\begin{equation}
V_{k}=A_{k}\otimes B_{k},  \label{V<A<B}
\end{equation}
In quantum optics experiments the $V_{k}$ can be computed from classical
optics. Using eq. (\ref{KrausRep}) we have 
\begin{eqnarray}
trO_{AB}\rho _{AB} &=&\sum_{k=1}^{n}trO_{AB}\left( V_{k}\rho
_{0AB}V_{k}^{\dagger }\right)   \nonumber \\
&=&\sum_{k=1}^{n}tr\left( V_{k}^{\dagger }O_{AB}V_{k}\right) \rho _{0AB}
\label{Kraus1}
\end{eqnarray}
the first line is referred to as the Schr\"{o}dinger picture and the second
line as the Heisenberg picture. Using our locality assumptions 
\begin{equation}
V_{k}^{\dagger }O_{AB}V_{k}=A_{k_{1}}^{\dagger }O_{A}A_{k_{1}}\otimes
B_{k_{2}}^{\dagger }O_{B}B_{k_{2}}\quad k=(k_{1},k_{2}).  \label{KrausLocal}
\end{equation}
We can now rewrite eq. (\ref{requirement}) as 
\begin{equation}
trO_{AB}^{(i,j)}\rho _{AB}=\sum_{u}p_{u}\left( trO_{A}^{(i)}\sigma
_{A}^{(u)}\right) \left( trO_{B}^{(j)}\sigma _{B}^{(u)}\right) .
\label{requirement1}
\end{equation}
If both $O_{A}^{(i)}$ and $O_{B}^{(j)}$ are positive operators then the
right hand side of eq. (\ref{requirement1}) is the sum of positive terms and
there can be no relative phase information about the subsystems extracted
from this expression. Then since any Hermitian operator can be written as a
difference of two positive operators, there is a stringent condition on the
type of experiment that can be simulate by classically correlated systems.

It will be useful to record a few well-known results for pure states. Let $%
\Psi $ be a pure state with Schmidt representation given by 
\begin{equation}
\Psi =\sum_{k=1}^{N}c_{k}|\phi _{k}\rangle \otimes |\chi _{k}\rangle 
\label{SchmidtDecomp}
\end{equation}
\cite{Peres}. Recall that the set $\{|c_{k}|^{2},k=1,\cdots ,N\}$ is a
probability distribution and the states $\{\phi _{k}\}$ and $\{\chi _{k}\}$
are orthonormal sets in $H_{A}$ and $H_{B},$ respectively. The state $\Psi $
is separable if $c_{k}=\delta _{k1},$ otherwise it is entangled. If $c_{k}=1/%
\sqrt{N},$ the state is said to be maximally entangled. For a maximally
entangled pure state $\Psi ^{(M)}$ a Kraus operator applied at one subsystem
can be replaced by an operator applied at the second subsystem, 
\begin{equation}
\left( A_{s}\otimes B_{r}\right) \Psi ^{(M)}=\left( 1_{A}\otimes
B_{r}A_{s}^{T}\right) \Psi ^{(M)},  \label{II0}
\end{equation}
where $A_{s}\phi _{j}=\sum_{k}\phi _{k}\left( A_{s}\right) _{kj}$ and $%
A_{s}^{T}\chi _{k}=\sum_{j}\chi _{j}\left( A_{s}\right) _{kj}.$ In
experiments it is often useful to place objects corresponding to Kraus
operators such as beam splitters or lenses in both laboratories A and B,
however, for maximally entangled states this is entirely equivalent putting
them all in one of the laboratories. This fact is exploited in remote state
preparation \cite{Gisin,Rubin1,Lvovsky}.

\section{Entanglement witnesses}

For finite dimensional systems the question of whether a given state,
represented by an $N\times N$ density matrix, is entangled can be answered
by operators, called entanglement witnesses \cite{H3,Terhal}, in the space
of Hermitian operators. Entanglement witnesses have a geometric
interpretation as hyperplanes that separate a given entangled state from the
separable states \cite{H3,Terhal,PittRub}. In this geometric picture, an
optimal entanglement witness is one which is tangent to the surface of the
separable states. Recently, the optimal entanglement witnesses have been
characterized in terms of a set of coordinated local measurements \cite
{Guhne,Pittenger}. An entanglement witness, $W,$ is defined for an entangled
state $\rho $ as a Hermitian operator such that 
\begin{equation}
tr(\rho W)<0,  \label{EntWit1}
\end{equation}
and for any separable density matrix $\tau $ 
\begin{equation}
tr(\tau W)\geq 0.  \label{EntWit2}
\end{equation}
The entanglement witness is said to be optimal if there exists a separable
density $\tau _{0}$ such that equality holds, that is, $W$ determines a
plane tangent to the separable states.

Consider the simplest case of two qubits \cite{Guhne,Pittenger}. We shall
denote the Pauli matrices by $\sigma _{\mu },$ $\mu =0,\cdots ,3$ where $%
\sigma _{0}$ is the $2\times 2$ identity matrix. Consider the state 
\begin{equation}
\rho =|\Phi _{+}\rangle \langle \Phi _{+}|  \label{Maxent}
\end{equation}
where $\Phi _{+}=\frac{1}{\sqrt{2}}\left( |00\rangle +|11\rangle \right) $
is one of the Bell states, $|0\rangle $ corresponds to the spin-up state,
and $|1\rangle $ the spin-down state in the basis in which the Pauli matrix $%
\sigma _{3}$ is diagonal. It has been shown that the optimal entanglement
witness for this state is 
\begin{equation}
W=\frac{2}{3}\left( \sigma _{0}\otimes \sigma _{0}-3\tau _{0}\right) 
\label{W1}
\end{equation}
where $\tau _{0}$ can be written in terms of local projection operators, the
factor of $2/3$ is a normalization. For $j=1,2,3$ define the projections 
\begin{equation}
P_{j}^{\pm }=\frac{1}{2}\left( \sigma _{0}\pm \sigma _{j}\right) 
\label{Proj1}
\end{equation}
and 
\begin{equation}
P_{j}^{ab}=P_{j}^{a}\otimes P_{j}^{b},  \label{Proj2}
\end{equation}
then 
\begin{mathletters}
\begin{equation}
\tau _{0}=\frac{1}{6}\left[ \left( P_{3}^{++}+P_{3}^{--}\right) +\left(
P_{1}^{++}+P_{1}^{--}\right) +\left( P_{2}^{+-}+P_{2}^{-+}\right) \right] .
\label{tau0}
\end{equation}
We see that $\tau _{0}$ can be written as the sum of three pairwise
commuting operators that are, however, locally non-commuting observables
corresponding to the measurement each spin along one of three orthogonal
axes. For a fixed experimental set-up, one of the pairs of operators in the
parentheses can be measured. For example, the measurement of the state in
eq. (\ref{Maxent}) along the $1$-axis can be reproduced using the separable
state 
\end{mathletters}
\begin{equation}
\tau =P_{1}^{++}\rho P_{1}^{++}+P_{1}^{--}\rho P_{1}^{--}=\frac{1}{2}\left(
P_{1}^{+}\otimes P_{1}^{+}+P_{1}^{-}\otimes P_{1}^{-}\right) .  \label{tau}
\end{equation}
Similarly, the other two measurements can be reproduced if the input state
is a (different) separable state. However the measurement $trW\rho <0$ can
not be reproduced by a separable state, $\tau ,$ since for all such state $%
trW\tau \geq 0$.

We can exploit this geometric picture further to show that there may be many
observables that will satisfy eq. ~(\ref{requirement1}). Let $\tau _{0}$ be
the nearest separable state to $\rho $ in the sense that 
\begin{equation}
||\rho -\tau ||=\sqrt{tr\left[ (\rho -\tau )^{2}\right] }\geq ||\rho -\tau
_{0}||  \label{HSDist}
\end{equation}
for all separable states $\tau .$ Then there is an optimal entanglement
witness for $\rho $ through $\tau _{0}.$ Suppose $\tau _{0}$ has rank
greater than one. Then the hyperplane of the entanglement witness intersects
the separable densities in, at least, a line containing $\tau _{0}$. For any
separable state $\tau $ on the line 
\begin{equation}
tr\left[ (\rho -\tau _{0})(\tau -\tau _{0})\right] =0.  \label{III0}
\end{equation}
Using the fact that the separable density matrices form a compact convex
set, we can write 
\begin{equation}
\tau -\tau _{0}=\sum_{u}a_{u}\sigma _{A}^{(u)}\otimes \sigma _{B}^{(u)}
\label{III1}
\end{equation}
where the $a_{u}$ are real but not necessarily positive and the $\sigma
_{A}^{(u)}\otimes \sigma _{B}^{(u)}$ are extreme points of the line.
Therefore, $\tau -\tau _{0}$ is an observable that corresponds to local
measurements with classical communications, consequently, when the system is
in the entangled state $\rho $ its measurement of this variable can be
simulated using the separable state $\tau _{0}.$

\section{SPDC Experiments}

The general theory of the type of experiment shown in fig. 1 has been given
in \cite{Rubin}. The source discussed in that paper was maximally entangled
in frequency and transverse wave number. It was pointed out that the source $%
\rho _{0}$ could be envisioned as entangled in transverse modes which are
not necessarily plane wave modes. In the present context these modes are
generated by the POVM's acting on the maximally entangled transverse modes
of the beams generated by the SPDC crystal. Specifically, suppose the pump
is a plane wave of angular frequency $\omega _{p}$ normally incident on a
crystal that is infinite in the transverse direction. The state of the down
converted biphoton is 
\begin{equation}
\Psi _{0}=\sum_{j}f_{j}a_{s}^{\dagger }(\omega _{j},{\bf p}%
_{j})a_{i}^{\dagger }(\omega _{p}-\omega _{j},-{\bf p}_{j})|0\rangle 
\label{SPDC}
\end{equation}
where $a_{r}^{\dagger }(\omega ,{\bf p})$ is the creation operator for a
photon at the output face of the crystal with polarization $r,$ frequency $%
\omega ,$ and transverse wave vector ${\bf p.}$ Then we can write the state
propagated to the detectors located at ${\bf r}_{1}$ and ${\bf r}_{2}$ 
\begin{eqnarray}
\Psi _{AB} &=&\sum_{j}g_{A}({\bf r}_{1},\omega _{j},{\bf p}_{j})g_{B}({\bf r}%
_{2},\omega _{p}-\omega _{j},-{\bf p}_{j})f_{j}a_{s}^{\dagger }(\omega _{j},%
{\bf p}_{j})a_{i}^{\dagger }(\omega _{p}-\omega _{j},-{\bf p}_{j})|0\rangle 
\nonumber \\
&=&\sum_{j}f_{j}a_{s}^{\dagger }({\bf r}_{1},\omega _{j},{\bf p}%
_{j})a_{i}^{\dagger }({\bf r}_{2},\omega _{p}-\omega _{j},-{\bf p}%
_{j})|0\rangle   \label{Output}
\end{eqnarray}
where the operators in second line are defined in the obvious way (cf. eq. (%
\ref{KrausLocal})). For example, 
\begin{equation}
a_{s}^{\dagger }({\bf r}_{1},\omega _{j},{\bf p}_{j})=g_{A}({\bf r}%
_{1},\omega _{j},{\bf p}_{j})a_{s}^{\dagger }(\omega _{j},{\bf p}_{j}).
\label{IV0}
\end{equation}
is the creation operator at detector one. The propagators $g$ are determined
by the linear optical systems in each arm of the experiment (see appendix 
\ref{Notation} and \cite{Rubin} for detailed calculations). They can be
expanded in terms of any convenient set of orthogonal transverse modes as is
usually done when the field propagates in a fiber. It may be useful to
recall that $g_{A}({\bf r}_{1},\omega _{j},{\bf p}_{j})$ is a transformation
from the transverse momentum ${\bf p}_{j}$ to the transverse position ${\bf r%
}_{1T}$ . The mode label $j$ and the longitudinal variable ${\bf r}_{1L}$
are parameters

It is the entanglement of the pair in their transverse coordinates that is
most important for image transfer. In many experiments the maximally
entangled state is a good approximation. As shown in eq. (\ref{II0}), it is
the maximal entanglement that makes it possible to put all the optics in one
arm. If the pump is not a plane wave $\Psi _{0}$ is no longer maximally
entangled in the transverse direction.

In the quantum imaging papers using SPDC one produces an image or
interference pattern in coincident counting. Since the SPDC has a large wave
number bandwidth even for a narrow frequency range there is no image or
interference pattern behind the object plane. Both of these effects can be
simulated by classically correlated sources. A second important feature of
the quantum case is the relation between the position of the aperture in one
arm and the detector array in the second arm. These relations depend on the
phase relation between the two-photon amplitudes or biphoton and, as we
shall see, can not be reproduced by classically correlated sources. This is
a specific example of the result discussed in connection with eq. (\ref
{requirement1}).

Starting with eq. (\ref{Output}) we can first ask what happens if we simply
detect all the photons at B discarding all the photons at A 
\begin{eqnarray}
\rho _{B} &=&tr_{A}|\Psi _{AB}\rangle \langle \Psi _{AB}|  \nonumber \\
&=&\sum_{jk}f_{jk}a_{i}^{\dagger }({\bf r}_{2},\omega _{p}-\omega _{j},-{\bf %
p}_{j})|0\rangle \langle 0|a_{i}({\bf r}_{2},\omega _{p}-\omega _{k},-{\bf p}
_{k})  \label{onePhoton}
\end{eqnarray}
where 
\begin{equation}
f_{jk}=f_{j}f_{k}^{*}\int d^{2}{\bf r}_{1T\ }g_{A}({\bf r}_{1},\omega _{j},%
{\bf p}_{j})g_{A}^{*}({\bf r}_{1},\omega _{k},{\bf p}_{k})  \label{fjk}
\end{equation}
and ${\bf r}_{1T}$ is the transverse component of ${\bf r}_{1}.$ As pointed
out in \cite{Bennink,Abouraddy}, if $g_{A}$ is unitary the $%
f_{ik}=|f_{j}|^{2}\delta _{ik}$ and the marginal density is the same as that
of the source. We have not put indices on the $g_{A}$ and $g_{B}$ because we
are concerned with a fixed experimental set-up. However, if we consider
different set-ups, as we would in testing Bell inequalities, then we need to
index them.

We now consider coincidence detection. For simplicity we take our detectors
to be polarization independent, point detectors oriented along the $\widehat{%
{\bf e}}_{L}$ direction. The electric field operator for such a detector
located at ${\bf r}$ is given by 
\begin{eqnarray}
E &=&E^{(+)}+E^{(-)}\quad E^{(-)}=E^{(+)\dagger }  \nonumber \\
E^{(+)}({\bf r},t) &=&\sum_{j,q}E_{j}a_{q}(\omega _{j},{\bf p}_{j})e^{i({\bf %
k}_{j}\cdot {\bf r}-\omega t)}\quad {\bf k}_{j}={\bf p}_{j}+\widehat{{\bf e}}%
_{L}\sqrt{\frac{\omega _{j}^{2}}{c^{2}}-p_{j}^{2}}  \label{detectorE}
\end{eqnarray}
where $E_{j}$ is determined by dimensional considerations, $q$ is the
polarization index, and $p_{j}<<\omega /c.$ The probability of two detectors
firing is proportional to 
\begin{equation}
tr\left( \rho _{AB}E_{2}^{(-)}E_{1}^{(-)}E_{1}^{(+)}E_{2}^{(+)}\right)
=|A_{12}|^{2}  \label{biphoton1}
\end{equation}
where 
\begin{eqnarray}
A_{12} &=&\langle 0|E_{1}^{(+)}E_{2}^{(+)}|\Psi _{AB}\rangle =\langle
0|E^{(+)}({\bf r}_{1},t_{1})E^{(+)}({\bf r}_{2},t_{2})|\Psi _{AB}\rangle  
\nonumber \\
&=&\sum_{j}g_{A}({\bf r}_{1},\omega _{j},{\bf p}_{j})g_{B}({\bf r}%
_{2},\omega _{p}-\omega _{j},-{\bf p}_{j})f_{j}E_{j}E_{h}e^{-i\omega
_{j}t_{1}\pi }e^{-i\omega _{h}t_{2}},  \label{biphoton2}
\end{eqnarray}
$\omega _{h}=\omega _{p}-\omega _{j},$ and ${\bf k}_{h}=-{\bf p}_{j}+%
\widehat{{\bf e}}_{L}\sqrt{\frac{\omega _{j}^{2}}{c^{2}}-p_{j}^{2}}.$ $A_{12}
$ is the biphoton amplitude.

In the imaging experiments one detector collects all the light that passes
through the object, $T$ in fig. 1, {\it i.e.} the correlation that is
measured is the correlation when the transverse component of ${\bf r}_{1}\ $%
is integrated over. To carry out the computation we assume degenerate phase
matching so that we can write $\omega _{1}=\omega _{p}/2+\nu $ and $\omega
_{2}=\omega _{p}/2+\nu ^{\prime }$ where frequency phase matching gives $\nu
^{\prime }=-\nu $. We require $|\nu |<<\omega _{p}/2.$ In the paraxial
approximation we can write 
\begin{equation}
A_{12}=\int d\nu u(\nu )e^{i\nu (t_{1}-t_{2})}e^{i\omega _{P}(t_{1}+t_{2})/2}%
\frac{1}{K^{2}}\int d^{2}xt({\bf x})e^{-iK^{2}({\bf x}_{2}-{\bf x})^{2}}
\label{ImageA12}
\end{equation}
where $t({\bf x})$ is the transmission function of $T$, and 
\begin{equation}
K=\left( \frac{2c}{\omega _{p}}\right) ^{2}\left( \frac{1}{d_{1}+d_{2}}+%
\frac{1}{d_{1}^{\prime }}-\frac{1}{F}\right) .  \label{K}
\end{equation}
The term containing $K$ comes from converting the sum over $j$ into an
integral in the amplitude $A_{12},$ and is therefore an effect of the
entanglement of the photons. When $K\rightarrow 0,$ the Gaussian becomes a
delta function and the coincident counting rate becomes proportional to $|t(%
{\bf x}_{2})|^{2}$. The fact that the image and object planes are related by
the geometrical optics condition $K=0$ was derived and experimentally
studied in \cite{Pittman1} where it was also generalized to the case of
non-degenerate SPDC.

We now ask whether this experiment can be simulated using a classically
correlated source of coherent states. Consider the classically correlated
state 
\begin{equation}
\tau _{AB}=\sum_{j}q_{j}\tau _{AB}^{(j)}.  \label{classical C1}
\end{equation}
where $\{q_{j},$ $j=1,2,\cdots \}$ is a probability distribution,

\begin{equation}
\tau _{AB}^{(j)}=\tau _{A}^{(j)}\otimes \tau _{B}^{(j)}\quad \tau
_{X}^{(j)}=|\psi _{X}^{(j)}\rangle \langle \psi _{X}^{(j)}|\quad X=A,B
\label{tauPure}
\end{equation}
and 
\begin{eqnarray}
|\psi _{A}^{(j)}\rangle  &=&\sum_{m}c_{m}^{(j)}|\{\alpha _{m}\}\rangle  
\nonumber \\
|\psi _{B}^{(j)}\rangle  &=&\sum_{m}d_{m}^{(j)}|\{\beta _{m}\}\rangle 
\label{coherentstates}
\end{eqnarray}
The state $|\alpha _{m}\rangle $ is a coherent state generated by the
creation operator $a_{s}^{\dagger }(\omega _{m},{\bf p}_{m})$ defined in eq.
(\ref{SPDC}) and the coherent states $|\beta _{m}\rangle $ are generated by
operators $a_{i}^{\dagger }(\omega _{p}-\omega _{m},-{\bf p}_{m})$. The
restriction to coherent states is not strictly necessary to show that the
results can be obtained from a classically correlated state, that is, we do
not have to assume that the states sent to A and B are themselves
''classical''.

For the experiment illustrated in fig. 2, it is not difficult to show that 
\begin{eqnarray}
C &=&\lim_{T\rightarrow \infty }\frac{1}{T}\int_{0}^{T}dt_{1}%
\int_{0}^{T}dt_{2}tr\left( \tau
_{AB}E_{2}^{(-)}E_{1}^{(-)}E_{1}^{(+)}E_{2}^{(+)}\right)
=\sum_{j}q_{j}tr_{A}\left( \tau _{A}^{(j)}E_{1}^{(-)}E_{1}^{(+)}\right)
tr_{B}\left( \tau _{A}^{(j)}E_{2}^{(-)}E_{2}^{(+)}\right)   \nonumber \\
&=&\int d^{2}p\int d^{2}p^{\prime }q({\bf p}{\bf ,}{\bf p}^{\prime })|g_{A}(%
{\bf r}_{1},\omega _{p}/2,{\bf p})|^{2}|g_{B}({\bf r}_{2},\omega _{p}/2,{\bf %
p}^{\prime })|^{2}  \label{incoherent imaging}
\end{eqnarray}
\cite{Bennink,Abouraddy}, where we have converted the discrete sum over
modes to integrals and defined 
\begin{equation}
\sum_{j}q_{j}|c_{m}^{(j)}|^{2}|d_{m^{\prime }}^{(j)}|^{2}|\alpha
_{m}|^{2}|\beta _{m^{\prime }}|^{2}\rightarrow d^{2}pd^{2}p^{\prime }q({\bf p%
}{\bf ,}{\bf p}^{\prime }).  \label{Limit}
\end{equation}
Equation (\ref{incoherent imaging}) is a special case of eq. (\ref
{requirement1}) and contains no relative phase information between A and B.
In arm B, the detector is placed in the focal plane of the second lens then,
as shown in the appendix eq. (\ref{gB2}), in the diffraction limited case 
\begin{equation}
|g_{B}({\bf r}_{2},\omega _{p}/2,{\bf p}^{\prime })|^{2}=G\delta ^{(2)}({\bf %
p}^{\prime }-\frac{\omega _{p}}{2c}\frac{1}{f_{2}}{\bf x}_{2})
\label{delta function}
\end{equation}
where ${\bf x}_{2}$ is the transverse coordinate of one of the point
detectors located at D2. Taking the correlation function 
\begin{equation}
q({\bf p,p}^{\prime })=\delta ^{(2)}(\varepsilon {\bf p}^{\prime }-{\bf p)}
\label{ClassicalCor}
\end{equation}
we get for the average number of coincident counts per pulse 
\begin{equation}
C=C_{0}|t(\varepsilon \frac{f_{1}}{f_{2}}{\bf x}_{2})|^{2}.
\label{CountingRate}
\end{equation}
Note the detector in arm B and the object $T$ are both in the focal plane of
the lens. This can be interpreted as a point by point mapping of the mask
onto the plane of the detectors at B. Equation (\ref{incoherent imaging})
can be viewed as incoherent imaging \cite{Bennink,Goodman}. The fact that a
point by point imaging is at the heart of this type of experiment was noted
some time ago in ref. \cite{Pittman1}.

The classically correlated source modelled here is a stochastic source that
is broadband in spatial frequencies. This condition is included to ensure
that there is no image in the $D1$ plane, as can be seen by setting $g_{B}=1$
in eq. (\ref{incoherent imaging}). The main characteristic of this source is
that whenever it emits a beam to $A$ with transverse wave number ${\bf p}$
it emits a beam to $B$ with wave number ${\bf p/\varepsilon }.$ The SPDC
source has the same type of correlation but is significantly different
because it imprints relative phase information on the beams in A and B which
influences the correlation measurements. The important phase information
exists in the transverse momentum dependent factors $\psi ({\bf p},X)$ in
eq. (\ref{gAimage}) and eq. (\ref{gB}). These phase factors disappear in eq.
(\ref{incoherent imaging}). This places stricter conditions on where the
images are located than in SPDC.

The ''ghost'' interference experiment in ref. \cite{Strekalov} displayed
remarkable features. A simplified diagram is shown if fig. 3. The first
important feature is that an interference pattern can be seen in the
coincidence counting but if single photons are detected in the plane of $D1$
no interference is seen because the SPDC is a broad band spatial frequency
source. The second feature is that the period of the interference pattern is
determined by the distance $z$ from $T$ back to the surface of the crystal
and then to $D2$. For the two slit experiment the coincident counting rate
was proportional to $\cos ^{2}(pd/2)$ where $d$ is the slit separation and $%
p=2\pi (x_{2}/z\lambda ),$ $x_{2}$ being the position of a detector in array 
$D2.$ A similar result holds for the single slit case. As the authors of
ref. \cite{Strekalov} point out: ''The interference pattern is the same as
one would observe on a screen in the lane of $D2,$ if $D1$ is replaced by a
point-like light source and the SPDC crystal by a reflecting mirror.'' This
is just the Klyshko picture \cite{Klyshko}.

Let us examine this in a bit more detail. The detector $D1$ is at the origin
of the plane at ''infinity'' so it detects the ${\bf p=0}$ transverse mode.
Since the SPDC is a broad band spatial source there are many different modes
impinging on the slits. When a photon is detected by one of the detectors in 
$D2$ one of these modes ${\bf p}^{\prime }$ is selected and we get
coincidences when $p^{\prime }\approx 2\pi n/d$ where $d$ is the distance
between the slits and $n$ is an integer. For a given ${\bf p}^{\prime }$ the
maximum of the interference pattern in the plane at $D1$ corresponds to $%
\theta \approx p^{\prime }/(\omega /c).$ Because of the broad bandwidth in $%
p^{\prime },$ this interference pattern is averaged over and is not visible
without the coincidence measurement.

The interference pattern observed in \cite{Strekalov} can also be simulated
using a similar classically correlated source as is shown in \ref{ghost}. 
As for the case discussed in \cite{Bennink}, not all the features can be
reproduced and the placement of the lenses $L1$ and $L2$ in fig. 2 play a
critical role. It is important to understand what about these experiments is
being simulated. In both cases the pattern, image or interference fringes,
and the fact that there is no image or interference if one looks behind the
object are simulated. The results relating the placement of detector $D2$
and the object $T$ can not be simulated because, as we saw above, because
this arises from phase information stored in the entangled system and not
separately in the subsystems $A$ and $B.$ As in the imaging case this phase
information is that in the factors $\psi ({\bf p},X)$ in eq. (\ref{ghostgA})
and eq. (\ref{ghostgB}). From eq. (\ref{classical C}) it is easy to see that
for the double slit experiment the fringes are determined by $p=2\pi %
(x_{2}/f_{2}\lambda ),$ that is the distance $z$ in the SPDC experiment has
been replaced by the focal length of the lens in arm B.

In a recent paper \cite{Gatti} it has been shown that for an entangled
source one can obtain both the image and its Fourier transform by only
varying the position of a lens and the detector in the arm of the experiment
that does not contain the aperture plane. This paper exploits the fact that
for maximally entangled states one can locate the optics in either arm of
the experiment. The authors explain their results by using entanglement,
path indistinguishability, and complementarity in the transverse momentum
and position. The complementarity argument must be made with care since it
means different things to different people. The spatial and momentum (wave
number) complementarity is a wave property and classical electromagnetic
fields are waves. After all, one can display the spatial position and
spatial frequencies of a source by simply changing the position of a lens.
The path indistinguishability along with the quantum rule about adding
amplitudes is more convincing since that invokes the particle properties of
the photon which distinguish the quantum and classical pictures of the
electromagnetic field. By using special cases of mixed state, the authors
obtain a special case of eq. (\ref{requirement1}) and reach the same
conclusions as those in this paper for the imaging case. However, they
conclude, for their mixed states that they can not reproduce the
interference fringes. This is not a general result, as we have shown.

We have shown that it is often possible to simulate some features of an
experiment using an entangled source by a classically correlated source.
Therefore, it is important when invoking entanglement as critical to an
experimental result that the possibility of such a simulation is eliminated.
Violation of Bell inequalities is a special case of using entanglement
witnesses as observables to show that entanglement is necessary to explain
some experimental measurements. Some quantum imaging experiments with SPDC
sources are examples in which it is possible to simulate some of the
measurements with classically correlated sources. In particular, the imaging
pattern can be reproduced in coincidence counting even when there is no
image or interference pattern visible behind the object. On the other hand,
it is impossible to reproduce the aspects of the quantum experiments that
depend on the relative phase information in the state of the entangled
subsystems. This is a consequence of eq. (\ref{requirement1}).

We have not answered the general question of what local observables can be
simulated by a classically correlated source for a given entangled source.
Finally, it should be noted that the converse problem of when a classically
correlated state can be used to study entanglement has been studied
extensively starting with ref. \cite{Popescu}.

\section{Acknowledgments}

This work was stimulated by a NASA-DoD workshop held a JPL in November 2002
organized by Jonathan Dowling. I wish to express my gratitude for helpful
comments from by colleagues at UMBC, particularly, Arthur O. Pittenger for
comments and suggestions with regard to sections II and III, and to Milena
D'Angelo and Yanhua Shih for discussions about the quantum imaging problem.
This work was supported in part by NSF.

\section{Appendix}

\subsection{Notation\label{Notation}}

We briefly review the general form of a coincidence counting experiment \cite
{Rubin}. The positive frequency part of the electric field at time $t$ at
the input of a detector located at ${\bf r=}z\widehat{{\bf e}}_{z}+{\bf x}$
is given by 
\[
E_{\beta }^{(+)}({\bf r},t)=\int \int d\omega d^{2}pe^{-i\omega t}E(\omega
)g\left( {\bf r},\omega ,{\bf p}\right) a_{\beta }\left( \omega ,{\bf p}%
\right) , 
\]
where $a_{\beta }\left( \omega ,{\bf p}\right) $ is the annihilation
operator at the source for a photon of angular frequency $\omega $,
transverse wave number ${\bf p}$, and polarization $\beta $. The unit vector 
$\widehat{{\bf e}}_{z}$ is the inward normal to the surface of the detector,
and ${\bf x}$ is the transverse coordinate on the face of the detector. $%
E(\omega )$ is required for dimensional reasons and is a slowly varying
function that can be taken to be constant in the experiments discussed in
this paper. Finally, $g\left( {\bf r},\omega ,{\bf p}\right) $ is the
optical transfer function or Green's function determined by the classical
linear optical system between the source and the detector. It may be worth
noting that classically this quantity connects electromagnetic fields that
live in real space-time while in the quantum mechanical context it is a
mapping of operators that operate in Fock space. This dual nature is the
source of many results in this paper and is also the source of confusion
about the ''photon''. It is important to recognize that linear superposition
in classical electromagnetic theory plays an essential role in determining $%
g.$ In the quantum mechanical case $g$ is best thought of as being
determined by the boundary conditions defining the modes of the system
independently of the state of the system. That is, the modes are the same
whether the system is in a one photon state or in a ''classical'' state for
which modes are excited by coherent states.

As an example of the computation of $g,$ we consider arm A of fig. 1. We use
the notation $t({\bf x})$ for the transfer function of the object where $%
{\bf x}$ is the transverse spatial vector in the plane of T. The thin lens
transformation is 
\begin{equation}
l({\bf x},f)=\psi (|{\bf x}|,-\frac{\omega }{c}\frac{1}{f})  \label{Lens}
\end{equation}
where $f$ is the focal length and ${\bf x}$ is the transverse coordinate in
the plane of the lens. Then, after integrating over the plane of D1, 
\begin{eqnarray}
g_{A}\left( {\bf r}_{1},\omega ,{\bf p}\right) &=&\int d^{2}x_{a}\int
d^{2}x_{f}\int d^{2}x_{s}t({\bf x}_{a})h_{\omega }({\bf x}_{a}{\bf -x}%
_{f},d_{1}^{\prime })\times  \label{gA} \\
&&l({\bf x}_{f},f)h_{\omega }({\bf x}_{f}{\bf -x}_{s},d_{1})e^{i{\bf p}{\bf %
\cdot x}_{s}}.  \nonumber
\end{eqnarray}
where ${\bf r}_{1}{\bf =}(d_{1}+d_{1}^{\prime })\widehat{{\bf e}}_{z}$, $%
{\bf x}_{s}$ is in the output plane of the source, and in the Fresnel
approximation 
\begin{eqnarray}
h_{\omega }({\bf x},d) &=&\left( \frac{-i\omega }{2\pi c}\right) \frac{%
e^{i(\omega /c)d}}{d}\psi (|{\bf x|,}\frac{\omega }{c}\frac{1}{d})  \label{h}
\\
\psi (x,\frac{\omega }{c}P) &=&e^{i\frac{1}{2}(\frac{\omega }{c})Px^{2}}.
\label{psiFresnel}
\end{eqnarray}
This notation is very convenient \cite{Rubin,Vander}. Evaluating $g_{A}$
gives 
\begin{eqnarray}
g_{A}\left( {\bf r}_{1},\omega ,{\bf p}\right) &=&\left( -i2\pi \frac{\omega 
}{c}\right) \frac{f}{f-d_{1}^{\prime }}e^{i\frac{\omega }{c}%
(d_{1}+d_{1}^{\prime })}\psi ({\bf p},-\frac{c}{\omega }D)\times  \nonumber
\\
&&\int d^{2}x_{a}t({\bf x}_{a})e^{i{\bf p\cdot x}_{a}f/(f-d_{1}^{\prime
})}\psi ({\bf x}_{a},-k\frac{1}{f-d_{1}^{\prime }}),  \label{gAimage}
\end{eqnarray}
where $D=d_{1}+\frac{d_{1}^{\prime }f}{f-d_{1}^{\prime }}.$

For arm B taking ${\bf r}_{2}={\bf x}+d_{2}\widehat{{\bf e}}_{z}$ and
performing the integration over ${\bf x}_{s}$ assuming that source has a
very large cross-section, we have for each plane wave input mode 
\begin{equation}
g_{B}({\bf r}_{2},\omega ,{\bf p})=\int d^{2}x_{s}h_{\omega }({\bf x-x}%
_{s},d_{2})e^{i{\bf p}{\bf \cdot x}_{s}}=e^{i(\omega /c)d_{2}}e^{i{\bf p}%
{\bf \cdot x}}\psi (|{\bf p}|,-\frac{c}{\omega }d_{2})  \label{gB}
\end{equation}
which is the Fresnel approximation to the plane wave $\exp (i\sqrt{(\frac{%
\omega }{c})^{2}-{\bf p}^{2}}d_{2}+{\bf p\cdot x}).$ Now consider what
happens if we introduce an aperture function in this expression.

\subsection{Ghost interference\label{ghost}}

Figure 3 shows a schematic of the ghost interference experiment \cite
{Strekalov}. In this case 
\begin{eqnarray}
g_{A}\left( (d_{1}+d_{1}^{\prime })\widehat{{\bf e}}_{z},\omega \right) 
&=&\int d^{2}x_{a}\int d^{2}x_{s}h_{\omega }(-{\bf x}_{a},d_{1}^{\prime })t(%
{\bf x}_{a})h_{\omega }({\bf x}_{a}{\bf -x}_{s},d_{1})e^{i{\bf p}{\bf \cdot x%
}_{s}}.  \nonumber \\
&=&\left( \frac{-i\omega }{2\pi c}\right) \frac{e^{i(\omega
/c)(d_{1}^{\prime }+d_{1})}}{d_{1}^{\prime }}\int d^{2}x_{a}\psi (|{\bf x}%
_{a}|,\frac{\omega }{c}\frac{1}{d_{1}^{\prime }})t({\bf x}_{a})e^{i{\bf p%
\cdot x}_{a}}\psi (|{\bf p}|,-\frac{c}{\omega }d_{1})  \label{ghostgA}
\end{eqnarray}
and 
\begin{equation}
g_{B}\left( d_{2}\widehat{{\bf e}}_{z},\omega ^{\prime },{\bf p}^{\prime
}\right) =e^{i(\omega ^{\prime }/c)d_{2}}e^{i{\bf p}^{\prime }{\bf \cdot x}%
_{2}}\psi (|{\bf p}^{\prime }|,-\frac{c}{\omega }d_{2}).  \label{ghostgB}
\end{equation}
In the degenerate case, $\omega =\omega ^{\prime }$ may be replaced by $%
\omega _{p}/2$ in the $\psi $-function and ${\bf p}^{\prime }=-{\bf p}$.
Then defining the biphoton wave function to be
\begin{equation}
\Psi =\int d^{2}p\int d\nu u(\nu )a_{s}^{\dagger }\left( \frac{\omega _{p}}{2%
}+\nu ,{\bf p}\right) a_{i}^{\dagger }\left( \frac{\omega _{p}}{2}-\nu ,-%
{\bf p}\right) |0\rangle ,  \label{Biphoton}
\end{equation}
the amplitude is given by 
\begin{eqnarray}
A_{12} &=&\frac{e^{-i\omega _{p}(\tau _{1}+\tau _{2})}/2}{d_{1}^{\prime }}%
\int d\nu u(\nu )e^{i\nu (\tau _{1}-\tau _{2})}A  \label{ghostA12} \\
A &=&A_{0}\int d^{2}x_{a}\psi (|{\bf x}_{a}|,\frac{\omega _{p}}{2c}\frac{1}{%
d_{1}^{\prime }})t({\bf x}_{a})\psi (|{\bf x}_{a}-{\bf x}_{2}|,\frac{\omega
_{p}}{2c}\frac{1}{d_{1}+d_{2}})  \nonumber \\
&=&A_{0}\int d^{2}x_{a}\psi (x_{a},\frac{\omega _{p}}{2c}\left( \frac{1}{%
d_{1}+d_{2}}+\frac{1}{d_{1}^{\prime }}\right) )t({\bf x}_{a})e^{-i(\omega
_{p}/2c)({\bf x}_{2}/d_{1}+d_{2})\cdot {\bf x}_{a}}\psi (|{\bf x}_{2}|,\frac{%
\omega _{p}}{2c}\frac{1}{d_{1}+d_{2}}).  \label{GhostA}
\end{eqnarray}
In the far field Fraunhofer approximation we may replace the $\psi ^{\prime
}s$ in the last line by $1$ 
\begin{equation}
A_{12}=A_{0}\widetilde{t}(\frac{\omega _{p}}{2c}\frac{{\bf x}_{2}}{%
d_{1}+d_{2}})  \label{ghost biphoton}
\end{equation}
where $\widetilde{t}({\bf p})$ is the Fourier transform of the aperture
transfer function.

In the classical source case we take arm A to be the same as in fig. 3$.$
Then from eq. (\ref{ghostgA}) we have 
\begin{equation}
|g_{A}\left( (d_{1}+d_{1}^{\prime })\widehat{{\bf e}}_{z},\omega ,{\bf p}%
\right) |^{2}=G_{A}|\int d^{2}x_{a}\psi (|{\bf x}_{a}|,\frac{\omega }{c}%
\frac{1}{d_{1}^{\prime }})t({\bf x}_{a})e^{i{\bf p\cdot x}_{a}}|^{2}
\label{gAClassInt}
\end{equation}
In arm B of the experiment we use the set up shown in fig. 2, 
\begin{equation}
g_{B}({\bf r}_{2},\omega ,{\bf p})=\int \int d^{2}x_{l}d^{2}x_{s}h_{\omega }(%
{\bf x}_{2}{\bf -x}_{l},f_{2})P({\bf x}_{l})l({\bf x}_{l},f_{2})h_{\omega }(%
{\bf x}_{l}{\bf -x}_{s},d_{2})e^{i{\bf p}{\bf \cdot x}_{s}}
\label{gBClassical}
\end{equation}
where we have include a finite aperture for the lens, $P(x_{l}).$ To make
the evaluation of the integral simple, we shall take 
\begin{equation}
P({\bf x}_{l})=e^{-|{\bf x}_{l}|^{2}/2A}.  \label{Pupil}
\end{equation}
Then 
\begin{equation}
|g_{B}({\bf r}_{2},\omega ,{\bf p})|^{2}=\left( \frac{\omega }{c}\frac{1}{%
f_{2}}\right) A^{2}e^{-({\bf p-}\frac{\omega }{c}\frac{1}{f_{2}}{\bf x}%
_{2})^{2}A}.  \label{gB1}
\end{equation}
In the limit of large $A,$ the exponential will be very small unless $({\bf %
p-}\frac{\omega }{c}\frac{1}{f_{2}}{\bf x}_{2})^{2}<<1/A.$ In this case we
may write to a good approximation 
\begin{equation}
|g_{B}({\bf r}_{2},\omega ,{\bf p}^{\prime })|^{2}=\left( \frac{\omega }{c}%
\frac{1}{f_{2}}\right) A\pi \delta ^{2}({\bf p}^{\prime }{\bf -}\frac{\omega 
}{c}\frac{1}{f_{2}}{\bf x}_{2}).  \label{gB2}
\end{equation}
Combining this result with eq. (\ref{gAClassInt}) in the far field limit, $%
d_{1}^{\prime }\rightarrow \infty $ in the integral, and the classical
correlation function eq. (\ref{ClassicalCor}) gives for eq. (\ref{incoherent
imaging}) 
\begin{equation}
C=C_{0}|\widetilde{t}(\varepsilon \frac{\omega _{p}}{2c}\frac{{\bf x}_{2}}{%
f_{2}})|^{2}  \label{classical C}
\end{equation}
where the lens L1 in fig. 2 is not necessary.

\subsection{Klyshko picture \label{KlyshkoPicture}}

Consider an experiment like that shown in fig. 1. For a biphoton source the
amplitude for the process can be written as 
\begin{equation}
A=\int d^{2}pg_{A}({\bf r}_{1},\omega ,{\bf p)}g_{B}({\bf r}_{2},\omega ,-%
{\bf p).}  \label{Klyshko1}
\end{equation}
The Klyshko picture is based on the observation that $g_{B}({\bf r}%
_{2},\omega ,-{\bf p)=}g_{B}({\bf r}_{2},-\omega ,{\bf p)}^{*},$ see for
example eq. (\ref{ghostgB}). The complex conjugate Green's function can be
viewed as a source located at ${\bf r}_{2}$ progagating backwards in time
reaching the SPDC crystal with transverse momentum ${\bf p.}$ At the crystal
the beam propagates forward in time to ${\bf r}_{1}.$

To generalize this picture, suppose that the beams with transverse momentum $%
{\bf p}$ have independent arbitrary phases so eq. (\ref{Klyshko1}) now takes
the form 
\begin{equation}
A=\int d^{2}pg_{A}({\bf r}_{1},\omega ,{\bf p)e}^{i\theta _{A}({\bf p}%
)}g_{B}({\bf r}_{2},\omega ,-{\bf p){\bf e}^{i\theta _{B}(-{\bf p})},}
\label{Klyshko2}
\end{equation}
where
\begin{eqnarray}
\langle {\bf e}^{i\theta _{C}({\bf p})}\rangle =0\quad \langle {\bf e}%
^{i\theta _{C}({\bf p})}{\bf e}^{-i\theta _{C}({\bf p}^{\prime 
})}\rangle & =&%
K\delta ^{(2)}({\bf p-p}^{\prime })\text{\quad }C=A\text{
or }B  \label{Klyshko3} \\
\langle {\bf e}^{i\theta _{A}({\bf p})}{\bf e}^{-i\theta _{B}({\bf p}%
^{\prime })}\rangle &=&0  \nonumber
\end{eqnarray}
and $K$ is a constant. With these assumptions 
\begin{equation}
\langle |A|^{2}\rangle =K\int d^{2}p|g_{A}({\bf r}%
_{1},\omega ,{\bf p)|}^{2}|g_{B}({\bf r}_{2},\omega ,-{\bf p)|}^{2}
\label{Klyshko4}
\end{equation}
which is of the form of eq. (\ref{incoherent imaging}) for the classical
correlation $q({\bf p,p}^{\prime })=K\delta ^{(2)}({\bf p}%
^{\prime }+{\bf p).}$

\pagebreak

\begin{figure}[t]
\centerline{
\input epsf
\setlength{\epsfxsize}{4in}
\epsffile{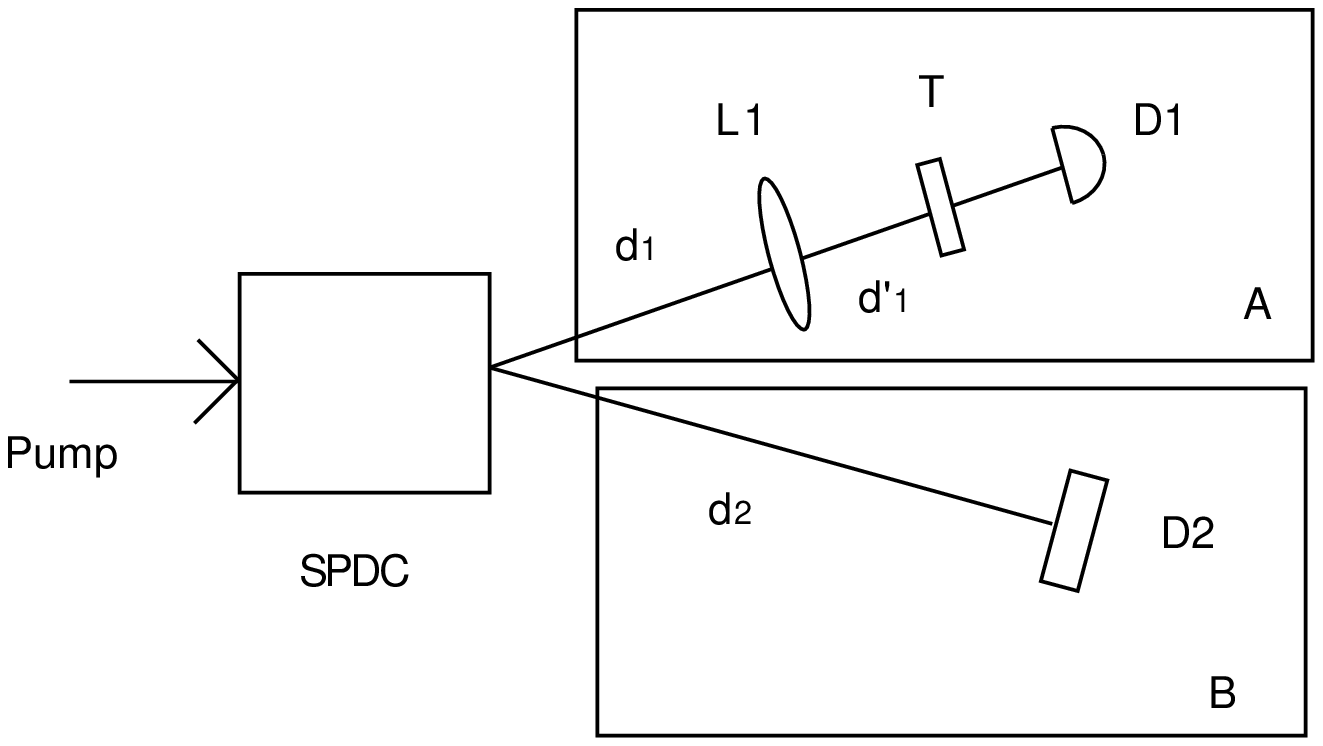}
}
\end{figure}

\begin{figure}[tbp]
\caption{Illustration of an SPDC experiment. In arm A there is a lens L1 a
distance $d_{1}$ from the output face of the SPDC crystal and an aperture T
a distance $d_{1}^{\prime}$ behind it. A fixed detector D1 that collects all
the light that is transmitted through T. In arm B the D2 is a planar array
of detectors located a distance $d_{2}$ from the crystal. The coincidence
circuit has been omited. }
\label{fig1.eps}
\end{figure}

\begin{figure}[t]
\centerline{
\input epsf
\setlength{\epsfxsize}{4in}
\epsffile{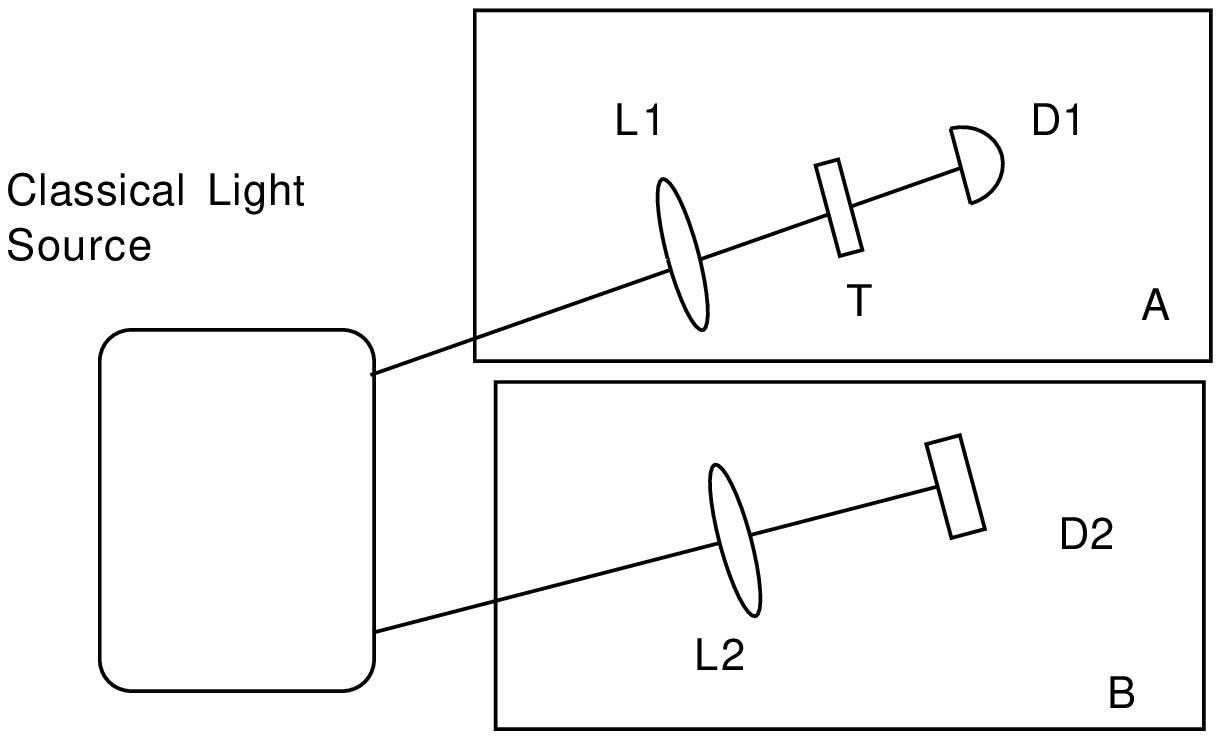}
}
\end{figure}
\begin{figure}[tbp]
\caption{Illustration of a classical analog of the experiment in fig.1. The
detectors are the same as those in fig. 1. The source is understood to be a
stocastic source that emits pairs of identical pulses into A and B. The
transverse wave number varies randomly from pulse pair to pulse pair. The
lenses can be thought of as part of the source or as part of the optical
systems in the two arms. However, it is important that T lies in the focal
plane of the lens L1 and the detector array plane lies in the focal plane of
L2. }
\label{fig2.eps}
\end{figure}

\begin{figure}[t]
\centerline{
\input epsf
\setlength{\epsfxsize}{4in}
\epsffile{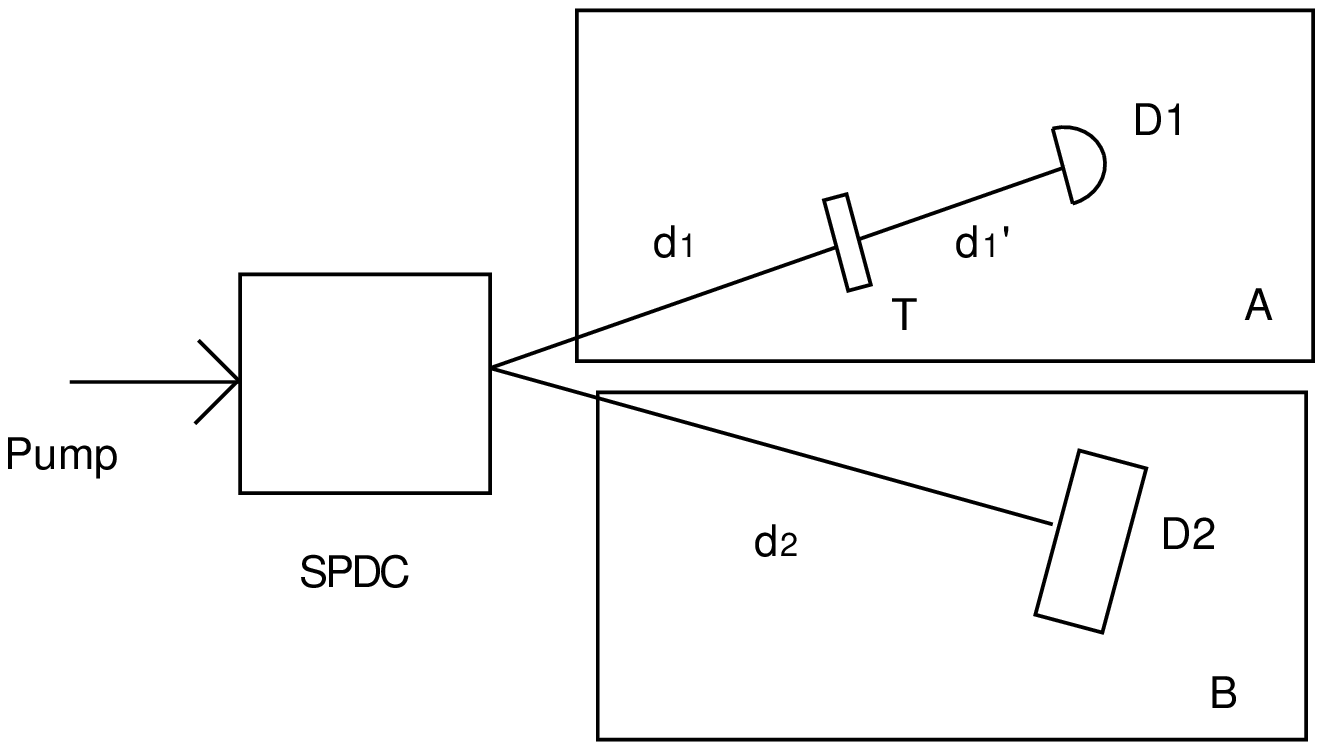}
}
\end{figure}

\begin{figure}[tbp]
\caption{ Illustration of the ghost interference experiment. The aperture T
is a double slit and the detector D1 is in the far field and only collects
light with zero transverse wave number.}
\label{fig3.eps}
\end{figure}


\begin{references}
\bibitem{convexset}  Recall that a set of objects $S$ is said to be convex
if for every pair of points $a,b\in S$ and any $t\in (0,1)$ the points $%
ta+(1-t)b$ are also in the set. That is to say, the straight line connecting
any two point in the set lies in the set.
\bibitem{random}  The completely random state has the property that
probability of projecting on to an arbitrary state is independent of the
state. In the case in the text, $tr\left( \rho D_{0}\right) =1/N$ for any
state $\rho .$
\end{references}
\end{document}